\documentclass[%
 reprint,superscriptaddress,amsmath,amssymb,aps,prl,floatfix
]{revtex4-2}

\usepackage[colorlinks=true,allcolors=blue]{hyperref}
\usepackage{cleveref}
\usepackage{graphicx}
\usepackage{xcolor}
\usepackage{xspace}
\usepackage{tabularx}
\usepackage{bbm}

\newcommand{\gw}{\ensuremath{GW}\xspace}
\newcommand{\scgw}{sc$GW$\xspace}

\newcommand{\aSn}{$\alpha$-Sn\xspace}

\begin{document}

\title{Symmetry-resolved topology from interacting Green's functions}

\author{Gaurav Harsha}
\affiliation{Department of Chemistry, University of Michigan, Ann Arbor, MI 48109, USA}
\author{Selina Dirnb\"ock}
\affiliation{Department of Physics, University of Michigan, Ann Arbor, MI 48109, USA}
\affiliation{Institute of Solid State Physics, TU Wien, 1040 Vienna, Austria}
\author{Emanuel Gull}
\affiliation{Department of Physics, University of Michigan, Ann Arbor, MI 48109, USA}
\affiliation{Department of Physics, University of Warsaw, Pasteura 1, 02-093 Warsaw, Poland}
\author{Vojtech Vlcek}
\affiliation{Department of Chemistry and Biochemistry, University of California, Santa Barbara, CA 93106, USA}
\affiliation{Materials Department, University of California, Santa Barbara, CA 93106, USA}
\author{Dominika Zgid}%
\affiliation{Department of Chemistry, University of Michigan, Ann Arbor, MI 48109, USA}
\affiliation{Department of Physics, University of Michigan, Ann Arbor, MI 48109, USA}
\affiliation{Department of Physics, University of Warsaw, Pasteura 1, 02-093 Warsaw, Poland}
\date{\today}

\begin{abstract}
Topological phases in narrow-gap materials often arise from a delicate interplay of spin--orbit coupling (SOC) and electron correlation, making their prediction sensitive to the underlying electronic-structure approximation. Electronic topology is commonly characterized using effective one-body Hamiltonians, but this approach becomes delicate near gap closings and does not reveal how a band inversion is realized in the interacting spectrum. Here we introduce a multi-messenger analysis that combines three complementary readouts of a single interacting Green's function: the Fu--Kane invariant of the topological Hamiltonian, the symmetry-resolved spectral function, and the symmetry-resolved orbital occupation. Together, they identify the topological class, resolve band connectivity, and track the orbital character of the inversion. We implement this framework within fully self-consistent relativistic \gw, treating SOC and electronic correlation on equal footing while eliminating dependence on a density-functional starting point. Applied to strain-tunable \aSn, the approach distinguishes its competing zero-gap, Dirac-semimetal, and topological-insulator regimes and provides an internally consistent description of their common band-inversion mechanism. More broadly, this framework offers a frequency-resolved route to diagnosing correlated topology across both gapped and near-critical states.
\end{abstract}

\maketitle

{\em Introduction:}
In weakly correlated materials, effective one-particle Hamiltonians provide established routes to topological invariants and surface spectra~\cite{bernevig_quantum_2006, fu_topological_3d_2007, fu_topological_inv_2007, murakami_phase_2007, bradlyn_topological_2017, cano_building_2018, cano_topology_2018, wu_wanniertools_2018, pizzi_wannier90_2020}. Many topological materials of interest, however, are narrow-gap systems in which the band inversion is governed by spin--orbit coupling (SOC) and electron correlation together, making predictions from mean-field or density-functional approaches sensitive to the chosen approximation~\cite{vidal_false-positive_2011, watkins_fidelity_2024}. Beyond-mean-field methods are therefore required~\cite{aguilera_gw_2013, aguilera_electronic_2015, nabok_bulk_2022}, with the interacting Green's function $G(k,\omega)$ often being the central object.

The established Green's-function route to topology reduces $G$ to the zero-frequency topological Hamiltonian $H_{\mathrm{top}}(k)=-G^{-1}(k,0)$. The invariant is then read from the symmetry or elementary-band-representation labels of its bands~\cite{wang_simplified_2012, deng_plutonium_2013, lessnich_elementary_2021}. This construction presumes a spectral gap and adiabatic continuity to a noninteracting reference. These conditions weaken in gapless and near-critical systems, where correlated topology is most delicate. Green's-function zeros, at which $-G^{-1}(k,0)$ diverges, can further mask a genuine inversion~\cite{misawa_zeros_2022, setty_electronic_2024}. A single index, moreover, labels a phase without exposing how the band inversion is realized.

In this letter, we go beyond this static characterization using three complementary readouts of one Green's function. The Fu--Kane topological-Hamiltonian invariant identifies the global class, the symmetry-resolved spectral function $A_\lambda(k,\omega)$ resolves band connectivity, and the symmetry-resolved orbital occupation $N_\lambda^{\mu}(k)$ tracks orbital character. Because all three derive from the same propagator, their agreement provides an internal consistency check, while their differences distinguish changes in global class, band connectivity, and orbital character.

We implement this analysis using our all-electron, fully self-consistent relativistic \gw method~\cite{yeh_relativistic_2022, yeh_fully_2022, abraham_relativistic_2024, iskakov_greenweakcoupling_2025} within the exact two-component (X2C) framework~\cite{kutzelnigg_quasirelativistic_2005, liu_quasirelativistic_2007, sun_exact_2009, saue_relativistic_2011, liu_essentials_2020}.
The \gw method treats orbital-selective screened interactions~\cite{hedin_new_1965, aryasetiawan_thegwmethod_1998, reining_gw_2018, golze_gw_2019}, scalar-relativistic effects, and SOC on equal footing. Full self-consistency eliminates dependence on a DFT starting functional~\cite{watkins_fidelity_2024, harsha_quasiparticle_2024}; hereafter \scgw denotes this framework.

Gray tin (\aSn) provides a stringent test because its near-zero gap and strong SOC place several competing phases within a narrow energy window. Small uniaxial strains drive its zero-gap ground state~\cite{groves_band_1963} into distinct topological regimes~\cite{hoffman_three-band_1989, xu_elemental_2017, rogalev_double_2017, gladczuk_study_2021, binda_spin-orbit_2021, anh_elemental_2021, chen_thickness-dependent_2022, zhang_large_2022, engel_growth_2024, polaczynski_3d_2024, alam_quantum_2024, li_chiral_2025}. Such strains are accessible in epitaxial films on InSb or CdTe~\cite{barfuss_elemental_2013, rogalev_double_2017, chen_thickness-dependent_2022, engel_growth_2024, polaczynski_3d_2024, alam_quantum_2024, rojas-sanchez_spin_2016, borlido_structural_2019, si_recent_2020, vail_gated_2021, jardine_first-principles_2023, bertoli_tuning_2025}, placing \aSn within reach of semiconductor and spintronic applications. Applying the three Green's-function readouts, we show that opposite [001] strains preserve a common band inversion but reorganize its connectivity into a tensile Dirac semimetal and a compressive bulk strong topological insulator.

\begin{figure*}[t]
    \centering
    \raisebox{-0.5\height}{\includegraphics[width=0.2\linewidth]{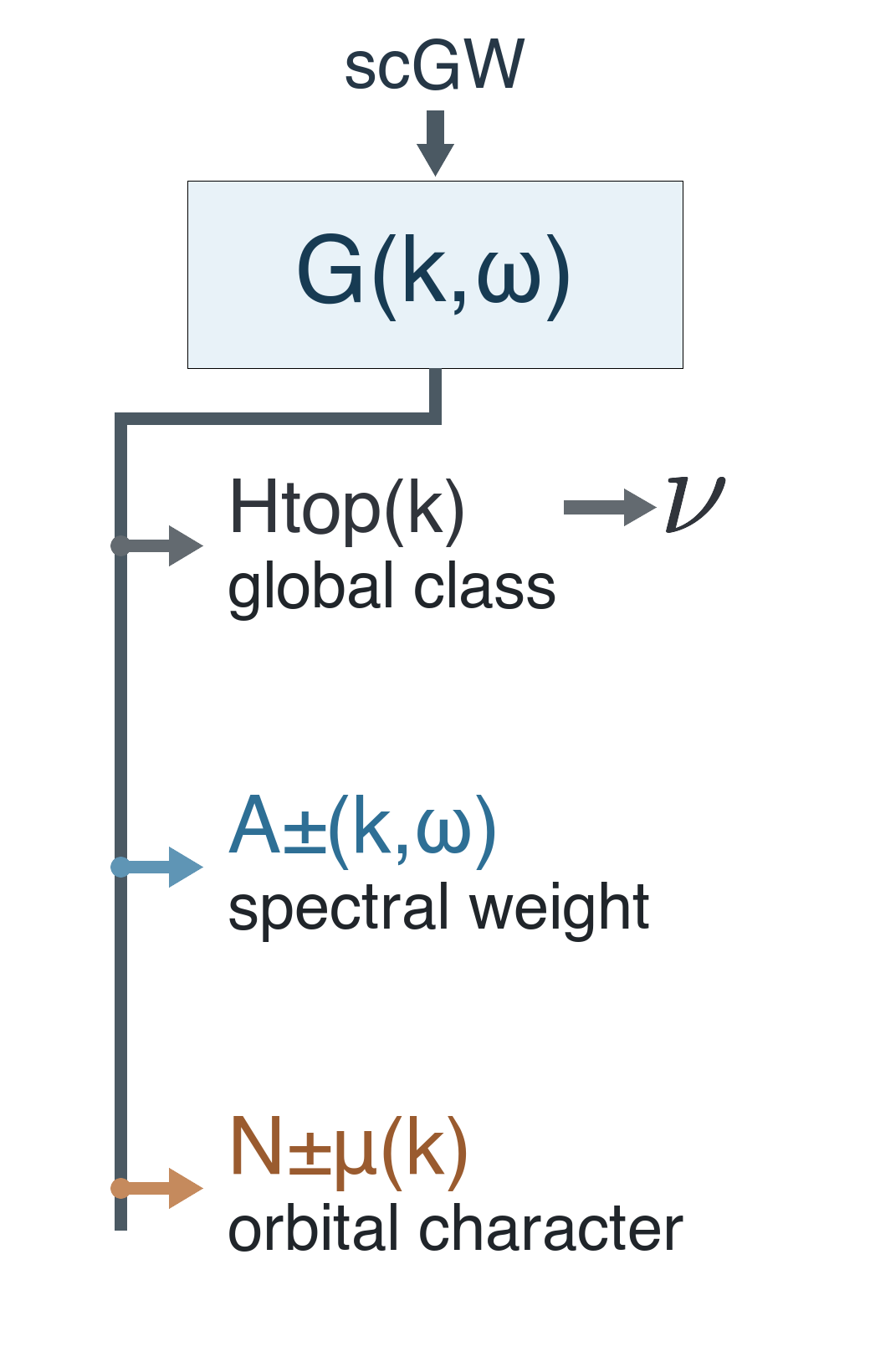}}
    \raisebox{-0.5\height}{\includegraphics[width=0.75\linewidth]{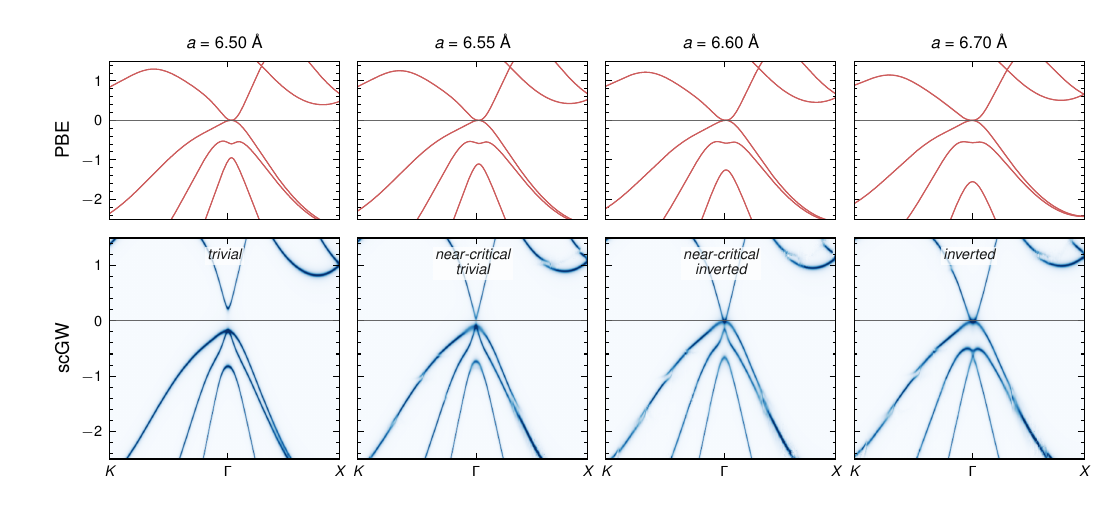}}
    \caption{\textbf{Green's-function framework and strain-driven topological evolution of \aSn.}
    Left: \scgw yields $G(k,\omega)$, whose topological Hamiltonian gives the global index $\nu$, while $A_\pm(k,\omega)$ resolves band connectivity and $N_\pm^\mu(k)$ tracks orbital character. Right: PBE bands (top) and \scgw spectra (bottom) along $K$–$\Gamma$–$X$ for $a=6.50$, $6.55$, $6.60$, and $6.70~\mathrm{\AA}$, referenced to $E_F$. PBE changes little across the series. In contrast, \scgw gives a trivial insulator at $6.50~\mathrm{\AA}$, retains a small trivial gap at $6.55~\mathrm{\AA}$, passes through a near-critical inverted state at $6.60~\mathrm{\AA}$, and reaches the inverted zero-gap semimetal at $6.70~\mathrm{\AA}$. The index changes from $\nu=0$ to $1$ across the inversion; for the gapless structures, $\nu=1$ denotes parity continuation.
    }
    \label{fig:bands}
\end{figure*}

{\em Green's-function framework:}
We characterize the topology of a converged \scgw Green's function through three complementary readouts of the same propagator. The global class follows from the topological Hamiltonian $H_{\mathrm{top}}(k)=-G^{-1}(k,i\omega\to0)$~\cite{wang_simplified_2012, deng_plutonium_2013, lessnich_elementary_2021}. At the eight time-reversal-invariant momenta $\Lambda_i$, the inversion eigenvalues $\xi_{2m}(\Lambda_i)$ of the occupied Kramers pairs give the Fu--Kane parity products $\delta_i$ and the strong $\mathbb{Z}_2$ index $\nu$:
\begin{equation}
    \delta_i=\prod_{m=1}^{N_{\mathrm{occ}}/2}\xi_{2m}(\Lambda_i),
    \qquad
    (-1)^\nu=\prod_{i=1}^{8}\delta_i.
    \label{eq:fu_kane}
\end{equation}
This index is rigorously defined for the gapped phases; for gapless structures, we use the parity product only as a continuation from the neighboring insulating phase.

To resolve how the inversion is realized, we project the full two-component correlated propagator onto complementary spatial-symmetry channels. For a spatial operation $O$ belonging to the little group of $k$ along the chosen high-symmetry line, we construct its orbital representation $D_{\mathrm{orb}}(O;k)$ and apply it identically to both components of the X2C basis. We restrict the two-channel construction to operations whose orbital representation satisfies $D_{\mathrm{orb}}(O;k)^2=e^{i\phi_O(k)}\mathbbm{1}$, where $\phi_O(k)$ is the Bloch phase associated with the lattice translation generated by $O^2$. After L\"owdin transformation to an orthonormal symmetrized-atomic-orbital (SAO) basis~\cite{lowdin_nonorthogonality_1970} and removal of any scalar nonsymmorphic Bloch phase, the normalized representation
\begin{equation}
    \begin{aligned}
        \widetilde D_{\mathrm{orb}}(O;k)
        &=e^{-i\phi_O(k)/2}\left[\mathbbm{1}_2\otimes D_{\mathrm{orb}}(O;k)\right], \\
        \widetilde D_{\mathrm{orb}}^2&=\mathbbm{1}.
    \end{aligned}
\end{equation}
defines the complementary projectors
\begin{subequations}
    \begin{align}
        P_\pm(k) &= \tfrac12[\mathbbm{1}\pm \widetilde D_{\mathrm{orb}}(O;k)], \\
        A_\pm(k,\omega) &= -\tfrac1\pi\,\Im\,\mathrm{Tr}[P_\pm(k)\,G(k,\omega+i\eta)], \\
        N_\pm^{\mu}(k) &= [P_\pm(k)\,\gamma(k)\,P_\pm(k)]_{\mu\mu},
    \end{align}
\end{subequations}
where $\gamma(k)$ is the correlated one-particle density matrix and $\mu$ indexes the SAOs; orbital-family occupations are obtained from $N_\pm^{\mu}(k)$.
Because $P_++P_-=\mathbbm{1}$ and $P_+P_-=0$, these projectors provide an exact and exhaustive decomposition of the interacting spectral function, $A(k,\omega)=A_+(k,\omega)+A_-(k,\omega)$, without requiring the projected operator to block-diagonalize the spinor Green's function. The $\pm$ labels therefore quantify the spatial-orbital symmetry character carried by the interacting spectrum; in the present construction they are not the eigenvalues of the complete double-group operation, and SOC may distribute a quasiparticle feature between both channels. We therefore refer to them below as spatial-orbital channels rather than exact double-group symmetry sectors. When the spin-off-diagonal components of the Green's function are small relative to the spin-diagonal terms, the spatial-orbital channels can provide an approximate description of band connectivity.

The symmetry-resolved spectral function $A_\pm(k,\omega)$ tracks band connectivity and the transfer of spectral weight between spatial-orbital channels, while the equal-time occupation $N_\pm^{\mu}(k)$ tracks the orbital character of the inversion. Both quantities are obtained directly from $G$, while the effective one-body topological Hamiltonian supplies the global index.

Computationally, we obtain the interacting Green's function using fully self-consistent \gw within the X2C framework, treating scalar-relativistic and SOC effects at the one-electron level. The PBE calculations and X2C Hamiltonian matrix elements are generated with \texttt{PySCF}~\cite{sun_pyscf_2018, sun_recent_2020}, while the \scgw iterations are performed with the \texttt{GREEN} package~\cite{iskakov_greenweakcoupling_2025}. The PBE and \scgw calculations use the all-electron \texttt{x2c-SVPall} Gaussian basis~\cite{pollak_segmented_2017} and a $6\times6\times6$ Monkhorst--Pack $k$-point mesh. The Green's functions are converged on the imaginary-time axis and analytically continued using the Nevanlinna method~\cite{fei_nevanlinna_2021} to obtain the spectral functions. Further details are provided in the End Matter.

{\em Results:}
To test the sensitivity to lattice scale and level of theory, we compare PBE~\cite{perdew_generalized_1996} and \scgw over $a=6.4$–$6.7~\mathrm{\AA}$, spanning published DFT estimates of $6.45$–$6.67~\mathrm{\AA}$~\cite{kufner_structural_2013,harsha_challenges_2024} and the experimental value $6.49~\mathrm{\AA}$~\cite{thewlis_thermal_1954, farrow_growth_1981}; \scgw predicts an equilibrium value near $6.45~\mathrm{\AA}$. Throughout this range, PBE yields a negative $\Gamma$–$L$ separation, with the conduction-band minimum at $L$ below the valence-band maximum at $\Gamma$, contrary to experiment~\cite{groves_band_1963, hoffman_three-band_1989} (SI Table I and Fig.~S1). Other semilocal functionals behave similarly~\cite{kufner_structural_2013, vlcek_improved_2015}, whereas HSE06~\cite{heyd_hybrid_2003} restores the correct ordering (SI Fig.~S5). By contrast, \scgw restores the ordering throughout the Brillouin zone and recovers the observed zero-gap semimetal at $a=6.7~\mathrm{\AA}$; we therefore use this structure as the electronic reference for the strain analysis.

Figure~\ref{fig:bands} contrasts the PBE and \scgw evolution along $K$–$\Gamma$–$X$. At $a=6.7~\mathrm{\AA}$, \scgw yields an ARPES-consistent zero-gap semimetal~\cite{engel_growth_2024} with an M-shaped valence band at $\Gamma$. Under compression, the near-$E_F$ spectrum evolves through a near-critical structure at $a\approx6.6~\mathrm{\AA}$; by $a=6.55~\mathrm{\AA}$, a local $\Gamma$-point gap of $0.15~\mathrm{eV}$ has opened and grows under further compression. The orbital character and spatial-symmetry character of this evolution are examined below using the symmetry-resolved diagnostics.

\begin{figure}[t]
    \centering
    \includegraphics[width=\linewidth]{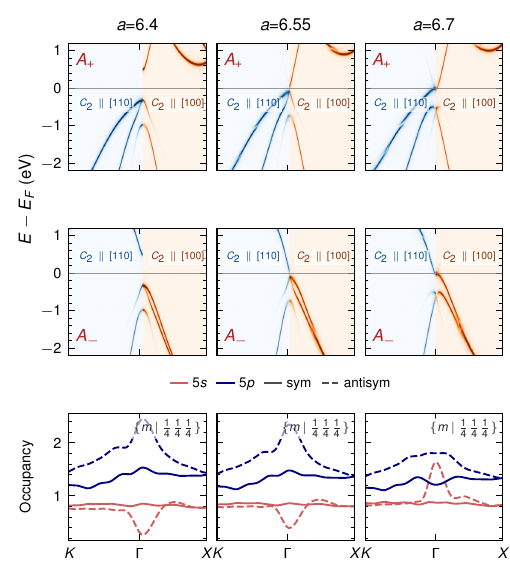}
    \caption{\textbf{Symmetry-resolved band inversion under uniform lattice variation.} Columns show trivial ($a=6.4~\mathrm{\AA}$), near-critical trivial ($a=6.55~\mathrm{\AA}$), and inverted ($a=6.7~\mathrm{\AA}$) \aSn. Top two rows: $C_2$-resolved \scgw\ spectral functions $A_\pm(k,\omega)$ near $E_F$; blue and orange denote the segment-specific $C_2$ projections on $K$–$\Gamma$ and $\Gamma$–$X$, respectively. At $a=6.55~\mathrm{\AA}$, the local $\Gamma$-point gap remains finite and the channel-resolved bands approach $E_F$ while retaining their trivial-phase character. At $a=6.7~\mathrm{\AA}$, the second-highest valence branch acquires weight near $\Gamma$ in $A_-$ along $K$–$\Gamma$ and in $A_+$ along $\Gamma$–$X$. Weak conduction-band weight appears in the complementary projections: $A_+$ along $K$–$\Gamma$ and $A_-$ along $\Gamma$–$X$. Bottom: glide-mirror--projected $5s$ (red) and $5p$ (blue) occupations; solid and dashed lines denote symmetric and antisymmetric combinations. The inverted occupation pattern becomes clear at $a=6.7~\mathrm{\AA}$.}
    \label{fig:diagnostic}
\end{figure}

We first validate the symmetry-resolved quantities against the known uniform-lattice inversion (Fig.~\ref{fig:diagnostic}). For \aSn, the spin-mixing components of the X2C Green's function are smaller than the dominant spin-diagonal contributions, allowing this spatial-orbital decomposition to provide a faithful and sensitive picture of the band connectivity. Here, the glide-resolved channels track the $\sigma$/$\sigma^{*}$ character of the inversion, while the $C_2$-resolved channels track changes in band connectivity. The reported $5s$ and $5p$ occupations are obtained by summing $N_\pm^{\mu}(k)$ over the SAOs in each respective orbital family. In the trivial $a=6.4~\mathrm{\AA}$ structure, the glide-projected antibonding $5s$ ($\sigma^*$) occupation is suppressed near $\Gamma$, while the $5p$ occupation peaks, matching isostructural silicon (SI Fig.~S7). By $a=6.7~\mathrm{\AA}$, the occupations exhibit the inverted $s$--$p$ ordering~\cite{rogalev_double_2017}. At $a=6.55~\mathrm{\AA}$, the channel-resolved bands approach $E_F$ while retaining the spatial-orbital character of the trivial $a=6.4~\mathrm{\AA}$ reference. At $a=6.7~\mathrm{\AA}$, spectral weight is redistributed onto the second-highest valence branch near $\Gamma$ in $A_-$ along $K$–$\Gamma$ and in $A_+$ along $\Gamma$–$X$. At the same time, a small amount of spectral weight appears in the conduction band just above $E_F$ near $\Gamma$ in the complementary projections: $A_+$ along $K$–$\Gamma$ and $A_-$ along $\Gamma$–$X$. This redistribution of spectral weight, together with the inverted $5s$/$5p$ occupation pattern, identifies the change in orbital ordering despite the absence of a bulk gap (full series in SI Fig.~S2). This known transition provides the control for the strained phases below.

The Fu--Kane parity product returns $\nu=0$, as expected, for isostructural silicon and for uniformly compressed \aSn ($a\lesssim6.55~\mathrm{\AA}$; Fig.~\ref{fig:bands}), certifying both as trivial gapped phases. Although the strong $\mathbb{Z}_2$ index is not rigorously defined for the gapless lattices, their fourfold states at $E_F$ share a common inversion parity, providing an unambiguous continuation from the neighboring inverted phase (full parity tables in the SI). { Across the uniform-lattice series, only $\delta_\Gamma$ changes sign between $a=6.55$ and $6.60~\mathrm{\AA}$, placing the latter on the inverted side of the transition; the gapless $6.60$ and $6.70~\mathrm{\AA}$ structures both carry the continued parity product $-1$ (SI Table II).} The symmetry-resolved spectrum and occupation extend occupation-based diagnostics~\cite{lopez_identifying_2025} by tracking how the connectivity and orbital character reorganize across the gapped and gapless structures.

PBE remains inverted throughout the studied range and responds much less strongly to compression than \scgw, whereas HSE06 reproduces the qualitative \scgw phase sequence, providing an independent check on the phase diagram (SI Fig.~S5).

\begin{figure}
    \centering
    \includegraphics[width=\linewidth]{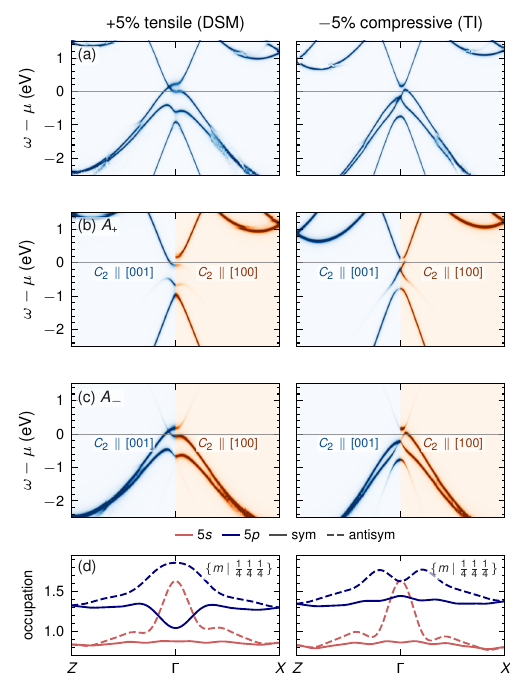}
    \caption{\textbf{Symmetry-resolved signatures of [001]-strained \aSn.} Columns compare $+5\%$ tension (3D Dirac semimetal) and $-5\%$ compression (strong topological insulator), referenced to $a=6.7~\mathrm{\AA}$. (a) Total \scgw\ spectrum $A(k,\omega)$ along $Z$–$\Gamma$–$X$. (b,c) Spatial-orbital channels $A_\pm(k,\omega)$, with the segment-specific $C_2$ axes labeled in each panel. (d) Glide-mirror--projected $5s$ and $5p$ occupations, resolved into symmetric and antisymmetric components. Both structures retain the inverted antibonding $5s$ ($\sigma^*$) occupation. Compression produces channel exchange and an occupied Dirac-like crossing within a gapped $\nu=1$ phase ($\Delta_\Gamma=0.36~\mathrm{eV}$); tension produces the symmetry-protected Dirac node, with $\nu=1$ denoting parity continuation.}
    \label{fig:strain_001_bands}
\end{figure}

We now apply this analysis to uniaxial [001] strain. Starting from the $a=6.7~\mathrm{\AA}$ reference, [001] strain splits the inverted state into two distinct regimes [Fig.~\ref{fig:strain_001_bands}(a)]. $c$-axis tension produces a 3D Dirac semimetal with nodes along $Z$–$\Gamma$, whereas compression folds the M-shaped valence feature into a twisted `X' at $\Gamma$ and produces a gapped bulk strong topological insulator. The Dirac-semimetal phase is already established at $+1\%$ tension, while under compression the bulk gap opens by $-1\%$ and the twisted-`X' feature develops progressively with increasing strain. We therefore show the $\pm5\%$ endpoints, where the characteristic dispersions are more clearly resolved (SI Fig.~S4). In epitaxial films, in-plane strain corresponds to $c$-axis strain of the opposite sign.

The glide-projected antibonding $5s$ ($\sigma^*$) occupation near $\Gamma$ remains essentially unchanged from the unperturbed lattice in both strained structures [Fig.~\ref{fig:strain_001_bands}(d); cf.\ Fig.~\ref{fig:diagnostic}], confirming that both inherit the inverted orbital character. For the gapless tensile phase, this indicates a common inverted-parity continuation rather than a rigorously defined insulating index. The distinction between the Dirac semimetal and topological insulator instead lies in their connectivity.

Under compression, the channels exchange character across the local $\Gamma$-point gap and along the adjoining $\Gamma$–$X$ segment: $A_+$ rises from the occupied, sub-$E_F$ crossing near $\Gamma$ into the conduction manifold, while $A_-$ descends from the conduction minimum into the valence manifold [Fig.~\ref{fig:strain_001_bands}(b,c)]. Under tension, the channels instead meet at the symmetry-protected Dirac node along $Z$–$\Gamma$.

Together, the inverted occupation and channel exchange identify the compressed phase as a strong topological insulator ($\nu=1$) with a local $\Gamma$-point gap of $0.36~\mathrm{eV}$. It also hosts a Dirac-like crossing approximately $0.2~\mathrm{eV}$ below $E_F$, whose spectral weight is carried predominantly by the $A_+$ channel and remains separated from the inverted $A_-$ band. Because this crossing lies in the occupied spectrum, it is directly accessible to ARPES; moderate hole doping could bring its near-linear carriers to $E_F$. Although a [001]-strained topological insulator has been established experimentally in relatively thick \aSn layers~\cite{engel_growth_2024}, its coexistence with this occupied Dirac-like crossing has not been reported.

{\em Discussion:}
The vanishing $\Gamma$-point gap of \aSn makes its electronic structure exceptionally sensitive to SOC and lattice distortions, posing a stringent challenge for semilocal DFT. Relativistic \scgw corrects the unphysical negative $\Gamma$–L separation, captures this lattice sensitivity, and supplies the correlated propagator underlying all three readouts.

We resolve a bulk topological insulator under [001] compression, with a Dirac-like band crossing just below the Fermi level, and a 3D Dirac semimetal under [001] tension. The bulk $\nu=1$ index of the compressed phase implies topological surface states at a time-reversal-preserving interface with a trivial insulator, although their dispersion is not determined here.
The compressed TI phase could be realized by growing \aSn films on the (001) surface of a diamond- or zinc-blende--type substrate with a lattice constant larger than that of gray tin.
Under uniform compression, by contrast, the inverted semimetal passes through a near-critical regime into a trivial insulator.
Together, uniform and [001] strain connect the zero-gap semimetal to trivial-insulator, Dirac-semimetal, and strong-topological-insulator regimes.

Independent HSE06 calculations reproduce the strain-induced band evolution, including the sub-$E_F$ linear dispersion of the compressed TI, showing that these features are not specific to \scgw.
The all-electron X2C \scgw treatment avoids DFT starting-point dependence and the pseudopotential approximation used in the hybrid calculations; a comparably rigorous all-electron relativistic hybrid treatment is not readily available and would have similar computational cost.

In \aSn, the combined readouts show that the same microscopic $s/p$ inversion can yield either a tensile Dirac semimetal or a compressive strong topological insulator, depending on band connectivity. Thus, band inversion alone does not determine whether the resulting phase is gapless or insulating.

The symmetry-resolved quantities are not new quantized invariants and must be interpreted relative to a trivial reference. They are informative when an inversion produces a resolvable change in symmetry character along a line possessing a suitable spatial symmetry. Symmetry operations of order greater than two can be accommodated by introducing additional channels. Because \aSn is weakly to moderately correlated and retains well-defined quasiparticles, an analysis based on Green's-function zeros is unnecessary here~\cite{gurarie_single-particle_2011, essin_bulk-boundary_2011, misawa_zeros_2022}. Whether the present framework remains effective when quasiparticle features become strongly incoherent or Green's-function zeros emerge remains to be explored, as does the assignment of a quantized invariant to a gapless correlated phase.

In summary, we have introduced a Green’s-function framework that combines the Fu–Kane invariant, symmetry-resolved spectral functions, and orbital occupations to distinguish global topology, band connectivity, and orbital character within a single interacting propagator. Applied to \aSn, fully self-consistent relativistic GW reveals that the same inverted orbital character can evolve into either a symmetry-protected Dirac semimetal under [001] tension or a strong topological insulator under [001] compression, demonstrating that orbital inversion alone does not determine the resulting connectivity or gap structure. The framework therefore provides a frequency-resolved route to characterizing correlated topological materials, including near-critical regimes where a static topological Hamiltonian gives an incomplete picture.

\begin{acknowledgments}
    This material is based upon work supported by the U.S. Department of Energy, Office of Science, Office of Advanced Scientific Computing Research and Office of Basic Energy Sciences, Scientific Discovery through Advanced Computing (SciDAC) program under Award Number DE-SC0022198.
    This research used resources of the National Energy Research Scientific Computing Center, a DOE Office of Science User Facility supported by the Office of Science of the U.S. Department of Energy under Contract No. DE-AC02-05CH11231 using NERSC award BES-ERCAP0029462 and BES-ERCAP0024293.
    S.D. acknowledges the Austrian Marshall Plan Foundation and TU Wien for enabling the research stay in Michigan.
    We also thank Kai Sun, Mariya Romanova, Aaron Engel, Christopher Palmstr{\o}m and Vibin Abraham for helpful discussions.
\end{acknowledgments}

\bibliography{biblio}

\clearpage
\section*{End Matter}

{\em Computational details:}
We employ fully self-consistent \gw (\scgw) to compute the interacting electronic structure of \aSn. In this approximation, the self-energy combines nonlocal exchange with dynamical correlation generated by the screened Coulomb interaction~\cite{hedin_new_1965}.
Relativistic effects are included within the exact two-component (X2C) framework~\cite{kutzelnigg_quasirelativistic_2005, liu_quasirelativistic_2007, sun_exact_2009, saue_relativistic_2011, liu_essentials_2020} for both PBE and \scgw.
In X2C, a unitary transformation decouples the positive- and negative-energy sectors of the four-component Dirac Hamiltonian. The resulting two-component electronic block, constructed with \texttt{PySCF}, includes scalar-relativistic and SOC effects. We retain the nonrelativistic Coulomb interaction and neglect two-electron relativistic corrections.
See Ref.~\cite{yeh_relativistic_2022} for further details on relativistic \gw theory.

All \scgw calculations are initialized from PBE DFT. In \scgw, the Green's function, screened interaction, and self-energy are updated to convergence, eliminating dependence on the initial DFT reference.
The PBE calculations and X2C Hamiltonian matrix elements are generated with \texttt{PySCF}~\cite{sun_pyscf_2018, sun_recent_2020}, while the \scgw iterations are performed with the \texttt{GREEN} package~\cite{iskakov_greenweakcoupling_2025}.
The PBE and \scgw calculations use the all-electron \texttt{x2c-SVPall} Gaussian basis~\cite{pollak_segmented_2017}.
For PBE and \scgw, Brillouin-zone integration uses a $6\times 6\times 6$ Monkhorst–Pack grid; the \scgw self-energy additionally includes finite-size corrections for the long-range Coulomb divergence~\cite{iskakov_greenweakcoupling_2025}.

The \scgw calculations use the imaginary-time formalism at inverse temperature $\beta = 1000~E_h^{-1}$; further increasing $\beta$ produces negligible changes in the spectra.
For analysis, a L\"owdin transformation maps the converged $G$ from the nonorthogonal atomic-orbital basis to an orthonormal symmetrized-atomic-orbital (SAO) representation~\cite{lowdin_nonorthogonality_1970}.
The spectral function is defined as
\begin{equation}
    A(k,\omega) = - \frac{1}{\pi} \Im \mathrm{Tr}\, G(k,\omega + i \eta).
    \label{eq:analyt-cont}
\end{equation}
We obtain $A(k,\omega)$ by Nevanlinna analytic continuation of the Matsubara Green's function~\cite{fei_nevanlinna_2021}; alternative continuation schemes are described in Refs.~\cite{fei_analytical_2021, huang_robust_2023, zhang_minimal_2024}.
The broadening is set to $\eta=0.5~\mathrm{m}E_h$ and controls the displayed spectral linewidth.

Along the tetragonal $Z$–$\Gamma$–$X$ path, the relevant $C_2$ axes differ: $C_2\|[001]$ on $Z$–$\Gamma$ and $C_2\|[100]$ on $\Gamma$–$X$. The glide mirror $\{m \, | \, \tfrac14\tfrac14\tfrac14\}$ is common to both segments and therefore provides a consistent $\sigma$/$\sigma^{*}$ resolution across the full path. The channel-exchange analysis is restricted to $\Gamma$–$X$, whose little group contains $C_{2x}$; the $\pm$ labels are those of the spatial-orbital projectors defined above.

Finally, the HSE06 calculations were performed using Quantum ESPRESSO v7.5~\cite{giannozzi_quantum_2020}, and include SOC within a plane-wave framework using norm-conserving two-component pseudopotentials~\cite{hamann_optimized_2013}. All Quantum ESPRESSO calculations use kinetic-energy cutoffs of $60~\mathrm{Ry}$ for the wavefunctions and $240~\mathrm{Ry}$ for the charge density and potential.

\end{document}

\title{Supplemental Information for ``Symmetry-resolved characterization of correlated topology from relativistic self-consistent \gw''}

\author{Gaurav Harsha}
\affiliation{Department of Chemistry, University of Michigan, Ann Arbor, MI 48109, USA}
\author{Selina Dirnb\"ock}
\affiliation{Department of Physics, University of Michigan, Ann Arbor, MI 48109, USA}
\affiliation{Institute of Solid State Physics, TU Wien, 1040 Vienna, Austria}
\author{Emanuel Gull}
\affiliation{Department of Physics, University of Michigan, Ann Arbor, MI 48109, USA}
\affiliation{Department of Physics, University of Warsaw, Pasteura 1, 02-093 Warsaw, Poland}
\author{Vojt\v{e}ch Vl\v{c}ek}
\affiliation{Department of Chemistry and Biochemistry, University of California, Santa Barbara, CA 93106, USA}
\affiliation{Materials Department, University of California, Santa Barbara, CA 93106, USA}
\author{Dominika Zgid}%
\affiliation{Department of Chemistry, University of Michigan, Ann Arbor, MI 48109, USA}
\affiliation{Department of Physics, University of Michigan, Ann Arbor, MI 48109, USA}
\affiliation{Department of Physics, University of Warsaw, Pasteura 1, 02-093 Warsaw, Poland}
\date{\today}

\maketitle

\section{PBE and \scgw band structures across the uniform-lattice series}
\begin{figure}[h]
    \centering
    \includegraphics[width=0.8\linewidth]{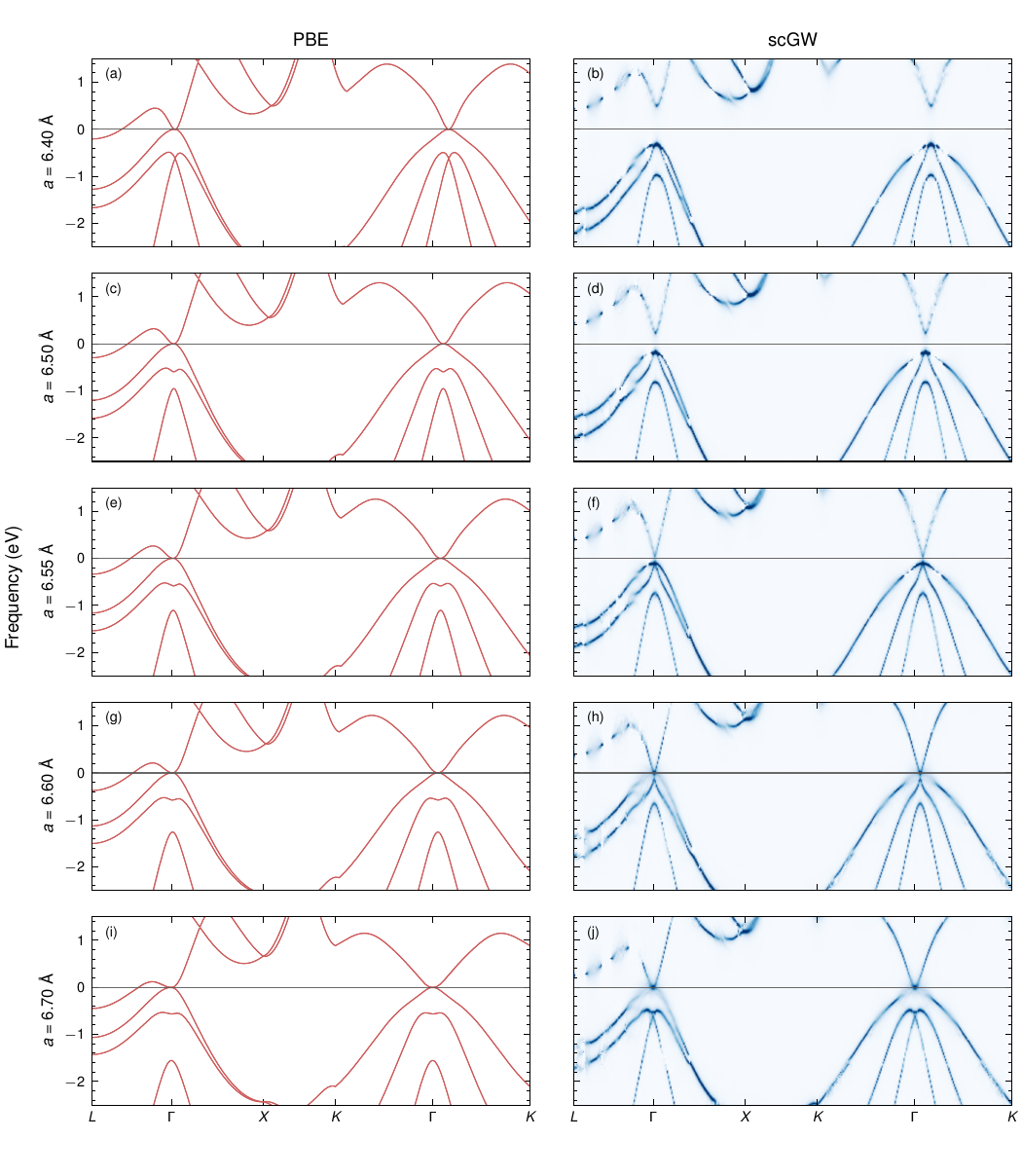}
    \caption{\textbf{Uniform-lattice electronic structure of \aSn.} PBE bands (left) and \scgw spectral functions $A(k,\omega)$ (right) along $L$--$\Gamma$--$X$--$K$--$\Gamma$--$K$ for lattice constants $a=6.40$--$6.70~\mathrm{\AA}$, referenced to $E_F$. The extended path displays both the local band evolution near $\Gamma$ and the relative position of the conduction minimum at $L$.}
    \label{fig:uniform_lgxkgk}
\end{figure}

Figure~\ref{fig:uniform_lgxkgk} compares the PBE bands and \scgw spectral functions over the full $L$--$\Gamma$--$X$--$K$--$\Gamma$--$K$ path. Including $L$ makes explicit the indirect $\Gamma$--$L$ overlap predicted by PBE and its correction by \scgw, while the vicinity of $\Gamma$ shows the evolution from the compressed trivial insulator through the critical structure to the inverted semimetal.

The corresponding local $\Gamma$ gaps and $\Gamma$--$L$ separations are collected in Table~\ref{tab:uniform_gaps}. The PBE $\Gamma$ point remains effectively gapless throughout the series while its increasingly negative $\Delta_{\Gamma L}$ signals a growing indirect overlap. In contrast, \scgw opens both separations under compression and drives the local $\Gamma$-point gap to zero near $a=6.60~\mathrm{\AA}$.

\begin{table}[h]
\centering
\caption{Uniform-lattice PBE and \scgw gap measures for \aSn, in eV. $\Delta_\Gamma=E_c(\Gamma)-E_v(\Gamma)$ is the local gap at $\Gamma$, and $\Delta_{\Gamma L}=E_c(L)-E_v(\Gamma)$ measures the indirect $\Gamma$--$L$ separation; a negative $\Delta_{\Gamma L}$ denotes band overlap. Values below the displayed precision appear as $0.00$.}
\label{tab:uniform_gaps}
\begin{tabular}{c cc cc}
\hline\hline
& \multicolumn{2}{c}{PBE} & \multicolumn{2}{c}{\scgw} \\
\cline{2-3}\cline{4-5}
$a$ (\AA) & $\Delta_\Gamma$ & $\Delta_{\Gamma L}$
          & $\Delta_\Gamma$ & $\Delta_{\Gamma L}$ \\
\hline
$6.40$ & $0.00$ & $-0.21$ & $0.82$ & $0.75$ \\
$6.50$ & $0.00$ & $-0.30$ & $0.39$ & $0.60$ \\
$6.55$ & $0.00$ & $-0.34$ & $0.15$ & $0.48$ \\
$6.60$ & $0.00$ & $-0.38$ & $0.00$ & $0.35$ \\
$6.70$ & $0.00$ & $-0.45$ & $0.00$ & $0.23$ \\
\hline\hline
\end{tabular}
\end{table}

\section{Interacting $\mathbb{Z}_2$ invariant from the correlated Green's function}
{ We evaluate the Fu--Kane parity products from the zero-frequency topological Hamiltonian constructed from the converged \scgw Green's function. Here we describe the construction and report the per-TRIM parity data. From the Dyson equation $G^{-1}(i\omega)=(i\omega+\mu)S-H_0-\Sigma_\infty-\Sigma_c(i\omega)$ we form the zero-frequency effective Hamiltonian $H_{\mathrm{qp}}=H_0+\Sigma_\infty+\Sigma_c(i\omega\!\to\!0)$, where $H_0$ is the one-body Hamiltonian, $\Sigma_\infty$ the static self-energy, and $\Sigma_c(i\omega\!\to\!0)$ the dynamical self-energy extrapolated to zero Matsubara frequency. In the nonorthogonal basis, this matrix is related to the main-text topological Hamiltonian by $H_{\mathrm{top}}=-G^{-1}(0)=H_{\mathrm{qp}}-\mu S$. The eigenvalues of this static Hamiltonian are not, in general, the physical excitation energies.}

{ We obtain the eigenstates of the static effective Hamiltonian by solving the generalized problem $H_{\mathrm{qp}}C=SCE$. At the eight time-reversal-invariant momenta (TRIM), the inversion parities $\xi=c^{\dagger}S\hat{P}c$ of the occupied Kramers pairs determine the products $\delta_i$ and hence the Fu--Kane index $\nu$ through $(-1)^{\nu}=\prod_{i=1}^{8}\delta_i$.}

We now give the working implementation in full, so that the per-TRIM $\delta_a$ are reproducible. The zero-frequency self-energy $\Sigma_c(i\omega\!\to\!0)$ is obtained from the imaginary-time self-energy $\Sigma_c(\tau)$ of the final scGW iteration: we transform $\Sigma_c(\tau)$ to the Matsubara axis on the intermediate-representation (IR) grid, take the Hermitian part, fit the four lowest positive Matsubara frequencies to the even form $\Sigma_c(i\omega)=a+b\,\omega^{2}$, and return the Hermitized constant $a$ as the static limit (the same recipe used to build the static quasiparticle Green's function). $H_0$ and the overlap $S$ are read on the full Brillouin-zone mesh from the mean-field input, while $\Sigma_\infty$ and $\Sigma_c(\tau)$ are read on the irreducible wedge and mapped to each TRIM by the stored symmetry table. { Assembling $H_{\mathrm{qp}}$ from these self-energy matrices avoids directly inverting the near-singular $G(i\omega\!\to\!0)$.} The inversion representative $\hat{P}$ is the space-group operation with point part $-\mathbbm{1}$, spinorized under the X2C treatment; at boundary TRIM its glide (fractional-translation) phase is removed by $\hat{P}\to e^{-i\psi/2}\hat{P}$ with $e^{i\psi}\mathbbm{1}=\hat{P}^{2}$, restoring a genuine $\pm1$ involution (at $\Gamma$, $\psi=0$, this is the identity).

At each TRIM we solve $H_{\mathrm{qp}}C=SCE$, order the states by $E$, and fill the lowest $N$ states, {where $N$ is the number of electrons per unit cell: $N=100$ for \aSn (two atoms with 50 electrons each) and $N=26$ for silicon with the GTH pseudopotential.} Rather than reading individual eigenstate parities---which are ill-defined inside the (near-)degenerate clusters of an all-electron spectrum, where the diagonalizer returns an arbitrary basis---we evaluate the trace of $\hat{P}$ over the occupied subspace, $\mathrm{Tr}\,M=\sum_{n\le N}c_n^{\dagger}S\hat{P}c_n=n_+-n_-$, where $n_\pm$ count the even/odd occupied states and $n_++n_-=N$, so $n_-=(N-\mathrm{Tr}\,M)/2$. { This trace is invariant under changes of basis within the selected occupied subspace, avoiding dependence on the arbitrary eigenvector basis within degenerate occupied shells.} Under the X2C spinor treatment Kramers degeneracy makes $n_-$ even, and the Fu--Kane per-pair product at that TRIM is $\delta_a=(-1)^{n_-/2}$. The distance of $\mathrm{Tr}\,M$ from the nearest integer, $\varepsilon=|\mathrm{Tr}\,M-\mathrm{round}(\mathrm{Tr}\,M)|$, measures whether the occupied subspace is closed under $\hat{P}$: a nonzero $\varepsilon$ signals a partially filled parity multiplet, i.e.\ an ill-defined (gapless) TRIM. We find $\varepsilon\le10^{-10}$ at every TRIM of every lattice reported here. This holds even at the gapless $\Gamma$ points of the $a\ge6.60~\mathrm{\AA}$ uniform lattices (Table~\ref{tab:parity_uniform}): the frontier Kramers pair immediately below $E_F$ remains a definite, parity-closed occupied set despite the vanishing bulk gap, so $\delta_\Gamma$ is unambiguous. Yet, the absence of a bulk gap makes the reported $\nu$ there a formal continuation rather than a quantized insulating invariant.

{ We present the uniform-lattice parity products and silicon control in Table~\ref{tab:parity_uniform}, the strain-series results in Table~\ref{tab:z2}, and the per-TRIM parities of the [001]-compressed topological insulator in Table~\ref{tab:parity}.}

{ Under uniform (hydrostatic) variation, the compressed lattices ($a\le6.55~\mathrm{\AA}$) are trivial insulators ($\nu=0$). The parity product changes from $+1$ at $a=6.55~\mathrm{\AA}$ to $-1$ at $6.60~\mathrm{\AA}$, bracketing the inversion between these sampled lattices. Table~\ref{tab:parity_uniform} shows that this change is carried entirely by $\delta_\Gamma:-1\to+1$, while the $L$ and $X$ parities remain fixed.} The expanded lattices are gapless at $\Gamma$, where the strong index is formal (no bulk gap) and the continuous symmetry-resolved diagnostic, rather than the invariant, characterizes the phase. { The silicon control in Table~\ref{tab:parity_uniform} returns $\nu=0$, as expected for a trivial insulator with the same diamond structure.}

\begin{table}[t]
\centering
\caption{{ Fu--Kane parity products and static $H_{\mathrm{qp}}$ gaps (eV) for cubic \aSn, with silicon as a trivial control. One entry represents each symmetry-equivalent TRIM class (multiplicities in parentheses), with $(-1)^\nu=\delta_\Gamma\delta_L^4\delta_X^3$. The sign change of $\delta_\Gamma$ brackets the inversion between $a=6.55$ and $6.60~\mathrm{\AA}$. For the gapless $a\ge6.60~\mathrm{\AA}$ lattices, $\nu$ denotes parity continuation rather than a quantized insulating invariant. The static gaps differ from the spectral gaps in Table~\ref{tab:uniform_gaps} and do not enter the parity product. The silicon control row reports its spectral $\Gamma$ gap and index only.}}
\label{tab:parity_uniform}
\begin{tabular}{c cc cc cc c c l}
\hline\hline
 & \multicolumn{2}{c}{$\Gamma$ ($\times1$)}
 & \multicolumn{2}{c}{$L$ ($\times4$)}
 & \multicolumn{2}{c}{$X$ ($\times3$)} & & & \\
\cline{2-3}\cline{4-5}\cline{6-7}
$a$ (\AA) & gap & $\delta_\Gamma$ & gap & $\delta_L$ & gap & $\delta_X$
          & $(-1)^\nu$ & $\nu$ & Character \\
\hline
$6.40$ & $0.93$ & $-1$ & $2.53$ & $+1$ & $5.13$ & $-1$ & $+1$ & $0$          & trivial insulator \\
$6.50$ & $0.46$ & $-1$ & $2.28$ & $+1$ & $4.99$ & $-1$ & $+1$ & $0$          & trivial insulator \\
$6.55$ & $0.20$ & $-1$ & $2.15$ & $+1$ & $4.92$ & $-1$ & $+1$ & $0$          & {near-critical trivial insulator} \\
$6.60$ & $0.00$ & $+1$ & $2.00$ & $+1$ & $4.83$ & $-1$ & $-1$ & $1^{\ast}$   & {near-critical inverted semimetal} \\
$6.70$ & $0.00$ & $+1$ & $1.77$ & $+1$ & $4.70$ & $-1$ & $-1$ & $1^{\ast}$   & {inverted zero-gap semimetal} \\
\hline
\multicolumn{7}{l}{{Silicon control: spectral $\Delta_\Gamma=4.60~\mathrm{eV}$}} & {$+1$} & {$0$} & {trivial insulator} \\
\hline
\multicolumn{10}{l}{$^{\ast}$ formal parity continuation ($\Delta_\Gamma=0$, no bulk gap).} \\
\hline\hline
\end{tabular}
\end{table}

{ Under uniaxial [001] strain (Table~\ref{tab:z2}), the} $\Gamma$ degeneracy is lifted and every case carries $\nu=1$; in particular the compressive flagship is a gapped bulk phase with $\Delta_\Gamma=0.36~\mathrm{eV}$ and $\nu=1$, establishing it as a true strong topological insulator.

\begin{table}[b]
\centering
\caption{Local $\Gamma$-point gaps and interacting strong $\mathbb{Z}_2$ index $\nu$ for {[001]-strained \aSn}. Strains are referenced to $a=6.7~\mathrm{\AA}$, and $\Delta_\Gamma=E_c(\Gamma)-E_v(\Gamma)$. { The \scgw gaps are extracted from spectral peak positions.} The finite $\Delta_\Gamma$ of the tensile structures does not constitute a bulk gap because symmetry-protected Dirac nodes remain along $Z$--$\Gamma$. Accordingly, $\nu$ is sharply defined for the gapped phases and denotes parity continuation for the Dirac semimetals.}
\label{tab:z2}
\begin{tabular}{l c c c l}
\hline\hline
System & $\Delta_\Gamma^{\mathrm{PBE}}$ (eV)
       & $\Delta_\Gamma^{\mathrm{sc}GW}$ (eV) & $\nu$ & Character \\
\hline
{[001]} $-5\%$ (squeeze) & $0.23$ & $0.36$ & $1$ & bulk TI \\
{[001]} $-1\%$ (squeeze) & $0.04$ & $0.08$ & $1$ & bulk TI \\
{[001]} $+1\%$ (stretch) & $0.03$ & $0.07$ & $1$ & Dirac semimetal \\
{[001]} $+5\%$ (stretch) & $0.15$ & $0.24$ & $1$ & Dirac semimetal \\
\hline\hline
\end{tabular}
\end{table}

{ For the $-5\%$ [001]-compressed structure referenced to $a=6.7~\mathrm{\AA}$, evaluating $H_{\mathrm{qp}}$} on the full $6\times6\times6$ mesh (30 irreducible points, covering the entire Brillouin zone by symmetry) confirms that the structure is gapped at every sampled $\mathbf{k}$. The minimum direct gap and the fundamental indirect gap of $H_{\mathrm{qp}}$ both occur at $\Gamma$ and equal $0.40~\mathrm{eV}$, with the conduction-band minimum lying above the valence-band maximum throughout the sampled zone. { The $0.40~\mathrm{eV}$ static gap is listed in Table~\ref{tab:parity}, while the corresponding spectral peak-to-peak gap of $0.36~\mathrm{eV}$ appears in Table~\ref{tab:z2}; the difference reflects the static-Hamiltonian and dynamical spectral definitions.} The eight per-TRIM parity products are collected in Table~\ref{tab:parity}; the three $X$-type points (the two in-plane $X$ and the [001] $Z$) are odd, so $\prod_a\delta_a=-1$ and $\nu=1$, establishing a strong topological insulator.

\begin{table}[t]
\centering
\caption{{ Per-TRIM Fu--Kane parity products and static $H_{\mathrm{qp}}$ gaps for $-5\%$ [001]-compressed \aSn ($a=6.7~\mathrm{\AA}$). Each $\delta_a$ is the product over occupied Kramers pairs, giving $\prod_a\delta_a=-1$ and $\nu=1$. Parent FCC labels distinguish two in-plane $X$ points and one axial $Z$; the four $L$ points remain equivalent. The static gaps do not enter the parity product.}}
\label{tab:parity}
\begin{tabular}{l c c c}
\hline\hline
TRIM & $\mathbf{k}$ (scaled) & gap (eV) & $\delta_a$ \\
\hline
$\Gamma$ & $(0,0,0)$                       & $0.40$ & $+1$ \\
$L$      & $(0,0,\tfrac12)$                & $1.90$ & $+1$ \\
$L$      & $(0,\tfrac12,0)$                & $1.90$ & $+1$ \\
$X$      & $(0,\tfrac12,\tfrac12)$         & $4.86$ & $-1$ \\
$L$      & $(\tfrac12,0,0)$                & $1.90$ & $+1$ \\
$X$      & $(\tfrac12,0,\tfrac12)$         & $4.86$ & $-1$ \\
$Z$      & $(\tfrac12,\tfrac12,0)$         & $4.83$ & $-1$ \\
$L$      & $(\tfrac12,\tfrac12,\tfrac12)$  & $1.90$ & $+1$ \\
\hline
\multicolumn{3}{l}{$\prod_a\delta_a=-1 \;\Rightarrow\; \nu=1$} & strong TI \\
\hline\hline
\end{tabular}
\end{table}

The invariant thus certifies the topology wherever a bulk gap exists (the trivial compressed insulator and the [001] bulk TI), while the gapless Dirac-semimetal phases, for which a quantized insulating invariant is undefined, are precisely the regime the continuous diagnostic is designed to resolve.
\color{black}

\section{Symmetry-resolved diagnostic across the uniform-lattice series}
\begin{figure}[h]
    \centering
    \includegraphics[width=0.85\linewidth]{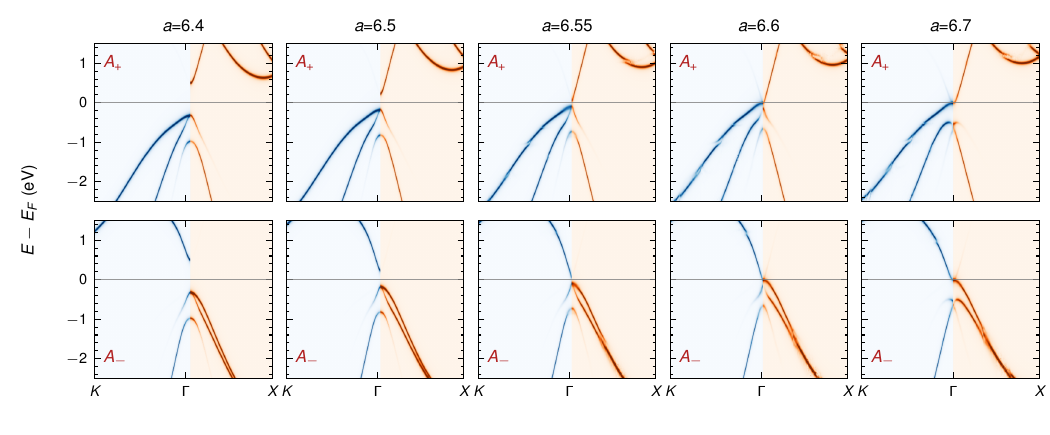}
    \caption{\textbf{Full uniform-lattice $C_2$ channel series of \aSn:} $C_2$-resolved \scgw spectral functions $A_+$ (top row) and $A_-$ (bottom row) near $E_F$ along $K$--$\Gamma$--$X$, for lattice constants $a=6.4$--$6.7~\mathrm{\AA}$ (columns). The per-segment $C_2$ operator is encoded by hue (blue on $K$--$\Gamma$, orange on $\Gamma$--$X$). { At $a=6.7~\mathrm{\AA}$, weight is redistributed onto the second-highest valence branch near $\Gamma$ in $A_-$ along $K$--$\Gamma$ and in $A_+$ along $\Gamma$--$X$, with weak conduction-band weight in the complementary projections.}}
    \label{fig:c2_series}
\end{figure}

Figure 2 of the main text shows the $C_2$-resolved \scgw spectral function $A_\pm(k,\omega)$ for three representative uniform lattices ({trivial, near-critical trivial, and inverted}). Figure~\ref{fig:c2_series} presents the full series, $a=6.4$--$6.7~\mathrm{\AA}$, along the $K$--$\Gamma$--$X$ high-symmetry path. The two spatial-orbital channels are drawn with the per-segment $C_2$ operator encoded by hue: sublattice-swapping $C_2$ on $K$--$\Gamma$ (blue) and sublattice-fixing $C_2$ on $\Gamma$--$X$ (orange). { As the lattice expands, the near-$E_F$ spectral weight is redistributed among the valence and conduction branches of the segment-specific $C_2$ projections through the near-critical regime, tracking the band inversion between $a=6.55$ and $6.60~\mathrm{\AA}$.}

\color{black}

\begin{figure}[h]
    \centering
    \includegraphics[width=0.6\linewidth]{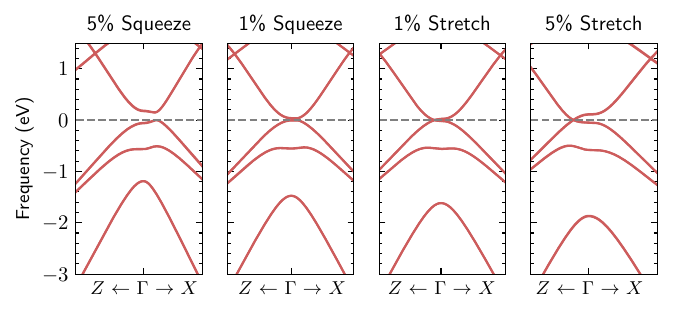}
    \caption{\textbf{Effect of [001] straining:} PBE bands for various degrees of compressive and tensile strain along the $c$-axis in \aSn. The results serve as a reference to compare the \scgw band structures in the main manuscript.}
    \label{fig:dft-001-bands}
\end{figure}

\section{PBE band structures for uniaxial strain along $c$-axis}
In the main manuscript, we analyze the effects of uniaxial straining along the $c$ axis, i.e., perpendicular to [001] plane, using the \scgw framework. We show that while a tensile strain induces a topological Dirac semimetal phase, compressive strain results in a topological insulator (TI). In addition to a finite local $\Gamma$-point gap, this TI phase exhibits a linearly-dispersing band crossing nearly 200 meV below the Fermi level.

{ We compare the PBE bands with the \scgw spectra under the same strain conditions.} Figure~\ref{fig:dft-001-bands} shows the PBE bands for the various $c$-axis strained \aSn lattices. For tensile strained configurations, PBE predicts a Dirac point along the $Z$-$\Gamma$ path, qualitatively consistent with \scgw. { Under compression, PBE opens a local $\Gamma$-point gap but does not reproduce the sub-$E_F$ crossing found in \scgw.} { This difference concerns PBE specifically; the HSE06 results below recover the crossing.}

\section{\scgw band structures for uniaxial [001] strain}
\begin{figure}[h]
    \centering
    \includegraphics[width=0.85\linewidth]{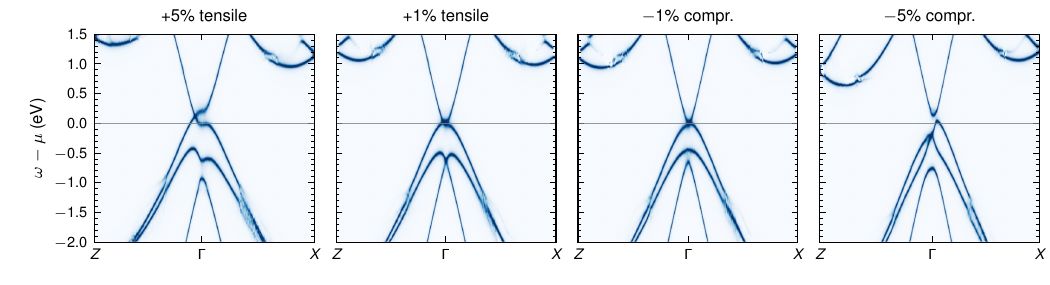}
    \caption{\textbf{[001]-strain series of \aSn:} \scgw spectral function $A(k,\omega)$ along $Z$--$\Gamma$--$X$ for $+5\%$ and $+1\%$ tension and $-1\%$ and $-5\%$ compression, referenced to $a=6.7~\mathrm{\AA}$.}
    \label{fig:strain_001_series}
\end{figure}

Figure~\ref{fig:strain_001_series} shows the \scgw spectral function $A(k,\omega)$ along $Z$--$\Gamma$--$X$ for a range of $c$-axis strains, from $+5\%$ tension to $-5\%$ compression, referenced to $a=6.7~\mathrm{\AA}$. The main text (Fig.~3) focuses on the $\pm5\%$ endpoints, which realize the 3D Dirac semimetal and the bulk topological insulator. The intermediate $\pm1\%$ cases shown here confirm that the Dirac point present under tension (on $Z$--$\Gamma$) evolves smoothly, and that the sub-$E_F$ linearly-dispersing crossing characteristic of the compressed topological insulator strengthens with increasing compression.

\color{black}

\begin{figure}[h]
    \centering
    \includegraphics[width=0.75\linewidth]{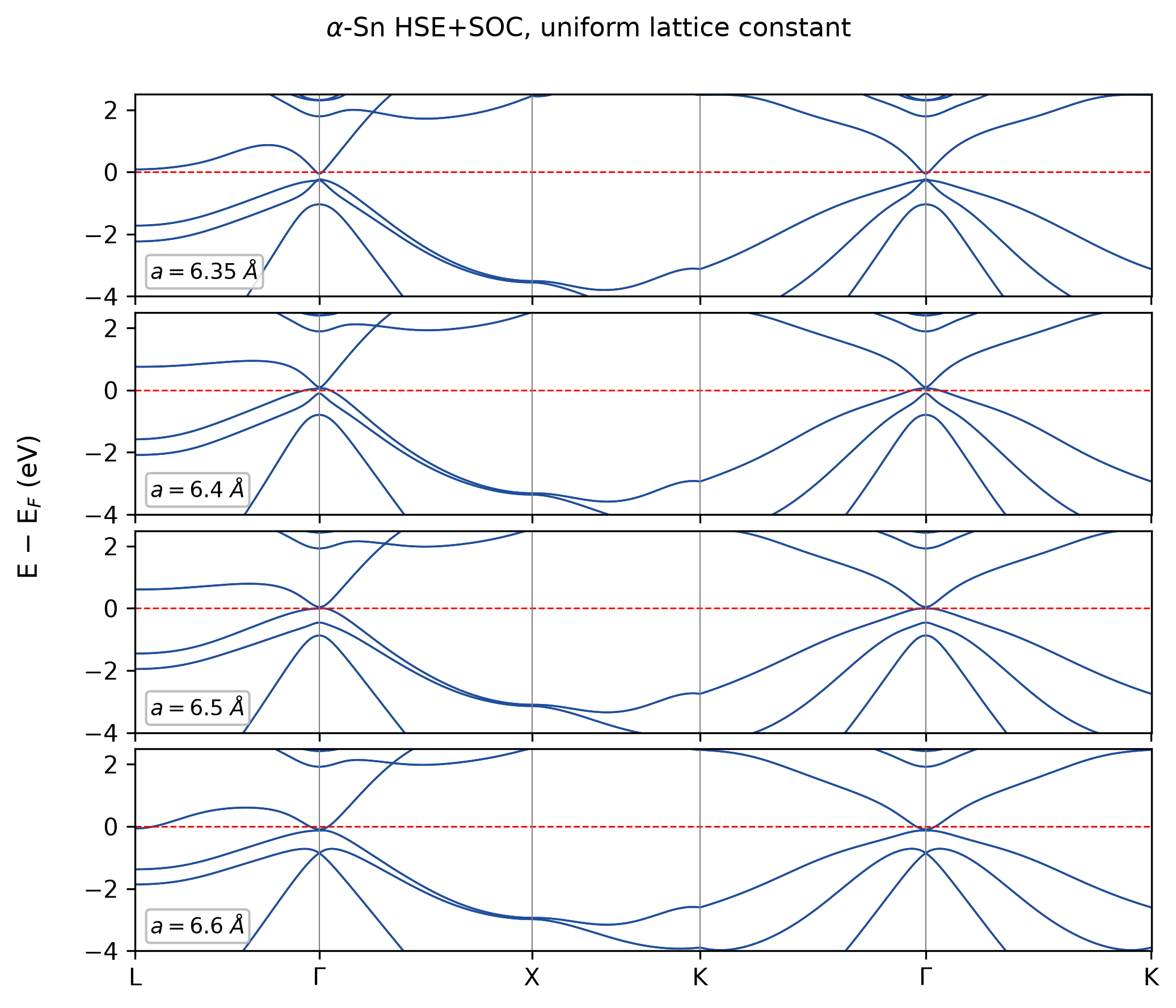}
    \caption{\textbf{HSE06 band structures of \aSn under uniform lattice variation.} HSE06 bands including spin--orbit coupling (SOC) along the $L$-$\Gamma$-$X$-$K$-$\Gamma$-$K$ path for lattice constants {$a=6.35$, $6.40$, $6.50$, and $6.60~\mathrm{\AA}$}, computed with a plane-wave, two-component-pseudopotential framework. HSE06 recovers the correct band ordering and reproduces the qualitative \scgw phase evolution.}
    \label{fig:hse_uniform}
\end{figure}

\section{HSE06 band structures for uniform lattice variation}
As an independent check on the \scgw phase diagram, Fig.~\ref{fig:hse_uniform} shows HSE06 band structures of \aSn, including SOC, for uniform lattice constants {$a=6.35$, $6.40$, $6.50$, and $6.60~\mathrm{\AA}$}, computed within a plane-wave, two-component-pseudopotential framework. { All Quantum ESPRESSO calculations reported here use kinetic-energy cutoffs of $60~\mathrm{Ry}$ for the wavefunctions and $240~\mathrm{Ry}$ for the charge density and potential.} In agreement with \scgw, and in contrast to PBE, HSE06 recovers the correct band ordering and reproduces the qualitative evolution {from an inverted zero-gap semimetal at the expanded end of the series toward a phase with a finite $\Gamma$-point gap upon compression.} { This agreement provides an independent check on the qualitative \scgw phase sequence.}
\color{black}

\section{HSE06 band structures for uniaxial [001] strain}
\begin{figure}[h]
    \centering
    \includegraphics[width=0.8\linewidth]{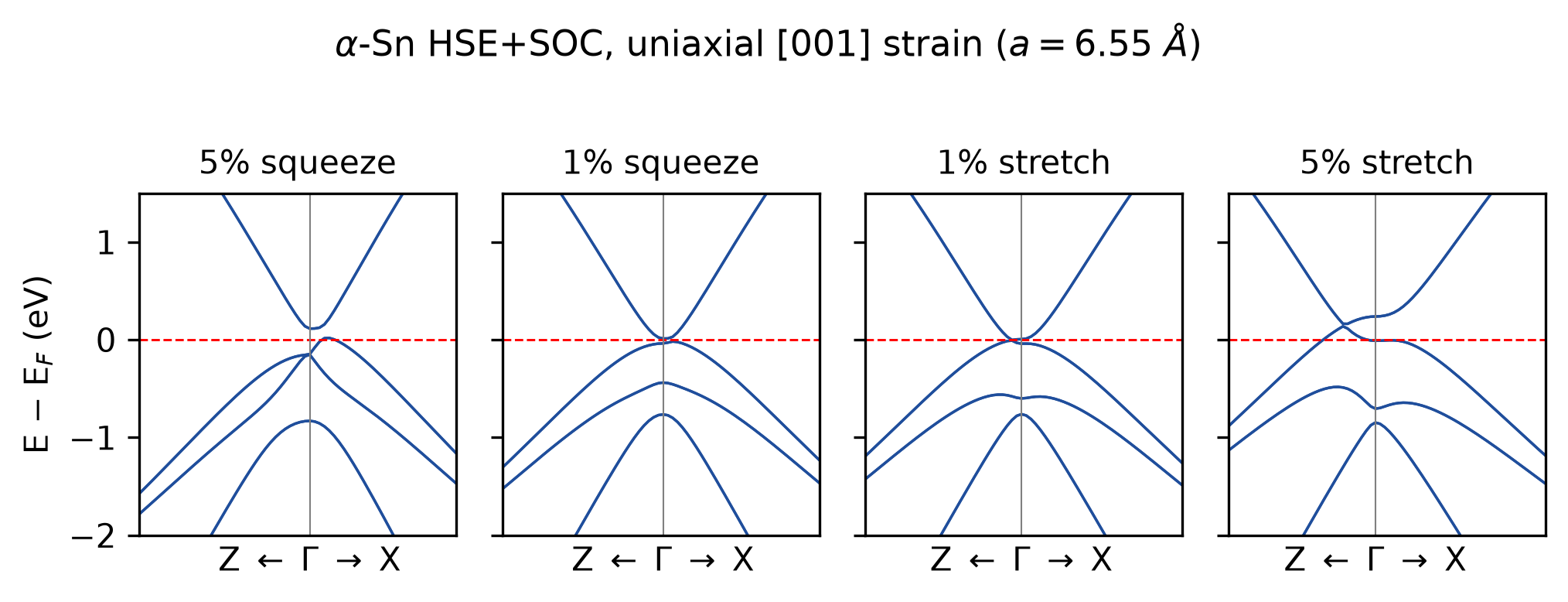}\\[4pt]
    \includegraphics[width=0.8\linewidth]{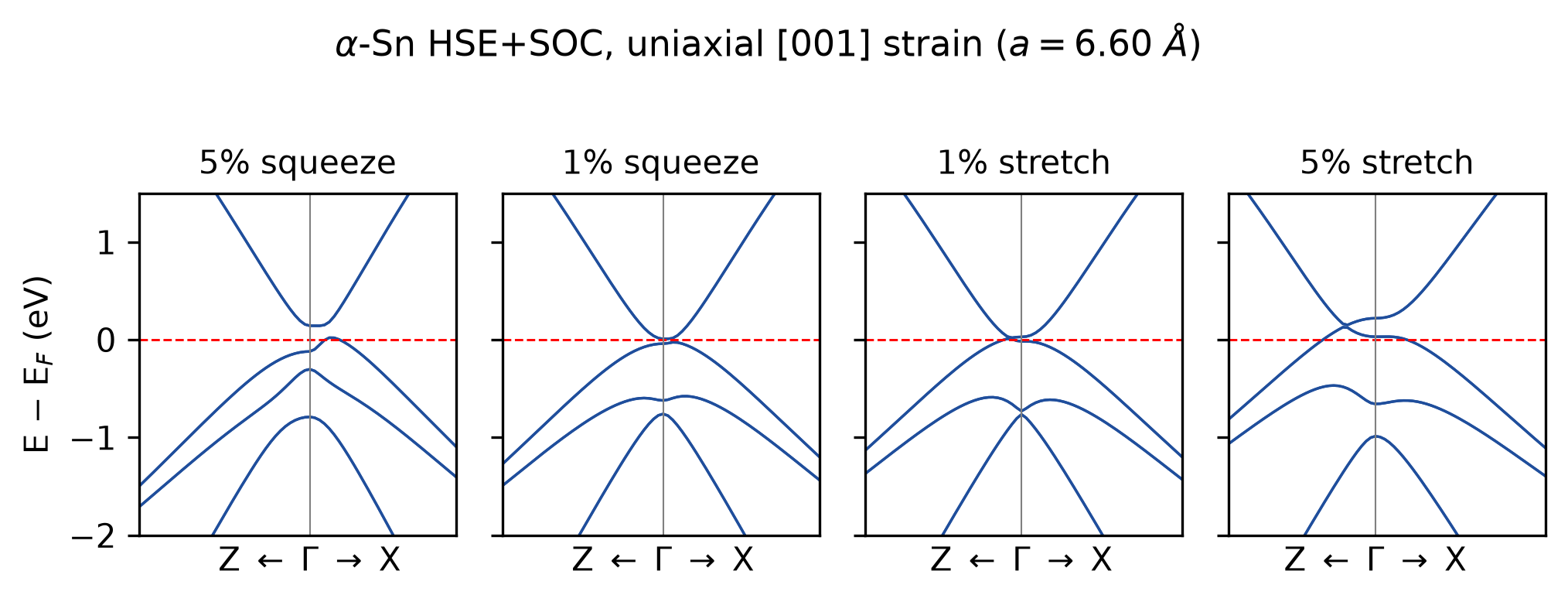}
    \caption{\textbf{HSE06 band structures of \aSn under uniaxial [001] strain.} HSE06$+$SOC bands along $Z$-$\Gamma$-$X$ for $5\%$ and $1\%$ compression and $1\%$ and $5\%$ tension, at reference lattice (a) $a=6.55~\mathrm{\AA}$ and (b) $a=6.60~\mathrm{\AA}$. Under compression, HSE06 reproduces the sub-$E_F$ linearly-dispersing crossing seen in \scgw.}
    \label{fig:hse_001}
\end{figure}

For the [001]-strained lattice, HSE06 also captures the sub-$E_F$ linearly-dispersing band crossing that characterizes the \scgw topological insulator. Figure~\ref{fig:hse_001} shows HSE06$+$SOC bands under $c$-axis strain at two nearby reference lattices, $a=6.55$ and $6.60~\mathrm{\AA}$; the below-$E_F$ crossing is most clearly resolved under compression near $a=6.55~\mathrm{\AA}$.

\color{black}

\section{Symmetry-resolved diagnostic for isostructural silicon (trivial control)}
\begin{figure}[h]
    \centering
    \includegraphics[width=0.8\linewidth]{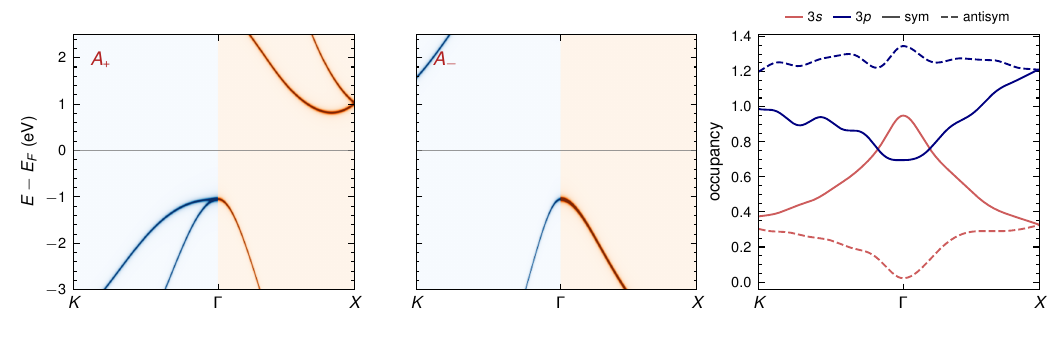}
    \caption{\textbf{Symmetry-resolved diagnostic for isostructural silicon (trivial control).} $C_2$-resolved \scgw spectral function $A_+$ (left) and $A_-$ (middle) near $E_F$ along $K$--$\Gamma$--$X$, with the per-segment $C_2$ operator encoded by hue (blue on $K$--$\Gamma$, orange on $\Gamma$--$X$); and (right) glide-mirror--projected $3s$ (red) and $3p$ (blue) occupation, split into symmetric (solid) and antisymmetric (dashed) combinations. The antibonding $3s$ ($\sigma^{*}$) occupation is suppressed near $\Gamma$, the signature of a trivial band insulator, in contrast to the inverted ordering in \aSn (main-text Fig.~2).}
    \label{fig:si_diagnostic}
\end{figure}

To confirm that the inverted {spatial-orbital channel} and occupation signatures identified in \aSn are genuine markers of band inversion rather than artifacts of the projection, we apply the identical symmetry-resolved diagnostic to isostructural silicon, a topologically trivial band insulator with the same diamond structure. Figure~\ref{fig:si_diagnostic} shows the $C_2$-resolved \scgw spectral function $A_\pm(k,\omega)$ and the glide-mirror--projected $3s$/$3p$ occupation along $K$--$\Gamma$--$X$. In contrast to \aSn, the antibonding $3s$ ($\sigma^{*}$) occupation (dashed red) collapses to nearly zero at $\Gamma$ ($\simeq 0.02$), while the bonding $3s$ ($\sigma$, solid red) is maximal, the fingerprint of a trivial, non-inverted band ordering. Neither {spatial-orbital channel} carries a near-$E_F$ crossing, consistent with the clean insulating gap. This trivial reference underlies the isostructural-silicon comparison drawn in the main text.

\color{black}